\documentclass[journal=jpcbfk,manuscript=letter,layout=twocolumn]{achemso}
\usepackage[utf8]{inputenc}
\usepackage{graphicx}
\usepackage{amsmath}
\usepackage{amssymb}
\usepackage{mathtools,cuted}
\usepackage{color}
\DeclareMathOperator{\Tr}{Tr}
\DeclareMathOperator{\FT}{FT}

\title{Interpretation of the THz-THz-Raman Spectrum of Bromoform}

\author{Ioan B. Magdău}
\affiliation{Division of Chemistry \& Chemical Engineering, California Institute of Technology, Pasadena, California 91125, United States}
\author{Griffin J. Mead}
\affiliation{Division of Chemistry \& Chemical Engineering, California Institute of Technology, Pasadena, California 91125, United States}
\author{Geoffrey A. Blake}
\email{gab@gps.caltech.edu}
\affiliation{Division of Chemistry \& Chemical Engineering, California Institute of Technology, Pasadena, California 91125, United States}
\affiliation{Division of Geological \& Planetary Sciences, California Institute of Technology, Pasadena, California 91125, United States}
\author{Thomas F. Miller III}
\email{tfm@caltech.edu}
\affiliation{Division of Chemistry \& Chemical Engineering, California Institute of Technology, Pasadena, California 91125, United States}

\begin{document}

\begin{abstract}
Nonlinear THz-THz-Raman (TTR) liquid spectroscopy offers new possibilities for studying and understanding condensed-phase chemical dynamics. Although TTR spectra carry rich information about the systems under study, the response is encoded in a three-point correlation function comprising of both dipole and polarizability elements. Theoretical methods are necessary for the interpretation of the experimental results. In this work, we study the liquid-phase dynamics of bromoform,  a  polarizable molecule with a strong TTR response. Previous work based on reduced density matrix (RDM) simulations suggests that unusually large multi-quanta dipole matrix elements are needed to understand the measured spectrum of bromoform. Here, we demonstrate that a self-consistent definition of the time coordinates with respect to the reference pulse leads to a simplified experimental spectrum.  Furthermore, we analytically derive a parametrization for the RDM model by integrating the dipole and polarizability elements to the 4$^{th}$ order in the normal modes, and we  enforce inversion symmetry in the calculations by numerically cancelling the components of the response that are even with respect to the field. The resulting analysis eliminates the need to invoke large multi-quanta dipole matrix elements to fit the experimental spectrum; instead, the experimental spectrum is recovered using  RDM simulations with dipole matrix parameters that are in agreement with independent \textit{ab initio} calculations.  The fundamental interpretation of the TTR signatures in terms of coupled intramolecular vibrational modes remains unchanged from the previous work.
\end{abstract}

\maketitle

\section{Introduction}

Two dimensional time resolved spectroscopy encompasses techniques in which a sequence of three or four coherent laser pulses is used to measure the response of the system with respect to the time delays.
The direct Fourier transform of this response corresponds to a frequency-frequency correlation function that carries rich information about the anharmonicity, coupling and non-linearity of the vibrational and electronic states. These characteristics elucidate the underlying dynamics and various energy dissipation processes, such as homogeneous and inhomogeneous broadening.

Each laser pulse in the sequence creates or destroys coherences via one of two physical processes: optical absorption (1$^{st}$ order in the field interaction) or Raman scattering (2$^{nd}$ order in the field interaction). These processes can take place in the visible (VIS) regime probing electronic states, the infrared (IR) regime measuring high frequency intra-molecular vibrational modes, or the terahertz (THz) regime probing low-frequency modes such as lattice phonons or inter-molecular vibrations. Combination of these various laser pulses leads to a range of different spectroscopic methods such as 2D-IR, 2D-VIS, VIS-IR, 2D-Raman, etc., each with its own applicability.

2D-IR spectroscopy was one of the first methods used to investigate the vibrational properties of liquids, and it has helped address  topics that include  structural fluctuations in water \cite{fecko2003ultrafast,zheng2007ultrafast,park2007hydrogen,ramasesha2013water}, proton shuttling in acidic solutions \cite{thamer2015ultrafast, dahms2017large} and protein dynamics \cite{shim2009two, bagchi2012ribonuclease, kratochvil2016instantaneous}. While 2D-IR has proven useful for studying high-frequency vibrational modes, there is strong motivation to extend multidimensional spectroscopy to the low-frequency regime. Important physical chemical processes, such as collective solvent motions \cite{heugen2006solute, Heisler2011, shalit2017terahertz} and dynamics of large biomolecules\cite{xu2006probing, Gonz2016, Turton2014} are driven by processes that takes place in the THz range.
The first method developed to study liquids in the THz domain was 2D-Raman spectroscopy\cite{Tokmakoff1997}. This method is 5$^{th}$ order in the laser field, and it has been shown to suffer from cascading effects where the 5$^{th}$ order response is plagued by contributions from higher intensity 3$^{rd}$ order processes\cite{Blank1999}. It is only with difficulty that these cascaded processes were overcome to yield the true 5$^{th}$ order 2D Raman response.\cite{Kaufman2002, Kubarych2002}

The advent of powerful THz sources \cite{Marder1989, Schneider2006} has enabled 3$^{rd}$ order hybrid spectroscopic techniques, in which one or two of the 2$^{nd}$ order Raman processes are replaced by 1$^{st}$ order THz absorption: RTT, TRT, and TTR \cite{lu2019two}. 
The RTT and TRT methods measure a THz emission and were first employed to study water and aqueous solutions \cite{savolainen2013two-dimensional,shalit2017terahertz,hamm2017perspective:,berger2019impact}. TTR builds upon the pulse-detection methods developed in THz-Kerr-effect spectroscopy,\cite{hoffmann2009terahertz,allodi2015nonlinear} which has been successfully applied in both polar and non-polar liquids \cite{sajadi2017transient, zalden2018molecular}.
The TTR approach was recently developed in our group \cite{finneran2016coherent,finneran20172d} and used to measure the 3$^{rd}$ order nonlinear response of bromoform \cite{hamm2019} (CHBr$_3$), carbon tetrachloride (CCl$_4$) and dibromodichloromethane (CBr$_2$Cl$_2$). These halogen liquids are ideal test systems for TTR because they exhibit  heavy intra-molecular modes in the THz frequency regime and  are Raman-active due to their strong polarizablity.\cite{teixeira1979raman} 

All three Raman-THz hybrid methods (i.e., RTT, TRT, and TTR) are complementary and measure different correlation functions involving the dipole and polarizability surfaces. These responses carry  rich information about the systems under study, including the nonlinearity of the dipole and polarizability surfaces and the anharmonicity and mechanical coupling of the various vibrational motions. However, this information is encoded in complex, three-point correlation functions that must be disentangled with the aid of  theoretical and computational methods \cite{steffen1996time,steffen1998population,okumura2003energy,hasegawa2006calculating,tanimura2009modeling,hamm2012two-dimensional-raman-terahertz,hamm2012note:,hamm20142d-raman-thz,ito2014calculating,ito2015notes,ikeda2015analysis,jo2016full,ito2016effects,ito2016simulating,finneran2016coherent,finneran20172d,sidler2019feynman}. This problem of interpretation is further complicated by the number of vibrational states that need to be considered.
Whereas infrared-active modes are usually in their vibrational ground state at room temperature and  typically only involve single excitations upon illumination, THz-active modes can be thermally excited at room temperature, and multiple transitions between the different states are possible.

In our previous work \cite{finneran20172d}, we developed a reduced density matrix (RDM) model to understand the TTR spectrum in liquid bromoform.
When the parameters for the nonlinear dipole elements in the RDM model were fit to best reproduce the experimental TTR signal, it was found that they assumed unexpectedly large values that did not agree with our accompanying \textit{ab initio} electronic structure calculations. Here, we reconcile this apparent inconsistency  by developing a more complete description of the TTR spectra from both the theoretical and experimental perspective.

\section{Method Development}

We begin by reviewing the RDM model of Ref.~\citenum{finneran2016coherent,finneran20172d}. In section Interpreting the Experimental Data (\ref{secIIa}), we describe the time-coordinate transformation necessary to reinterpret the experimental spectrum. In sections \ref{secIIb} and \ref{secIIc}, we explain the development of a new RDM model that fully accounts for the liquid symmetry and the symmetries of the dipole and polarizability matrices to  4$^{th}$ order in the normal modes.

\renewcommand{\thefootnote}{\roman{footnote}}

In RDM, we propagate the reduced density matrix $\rho(t)$ for a single bromoform molecule by solving the Liouville–von Neumann equation \cite{hamm2011concepts},
\begin{equation}
\label{equLvN}
\frac{\partial{\rho(t)}}{\partial t} = -\frac{i}{\hbar}\left[H(t),\rho(t)\right] -\Gamma \rho(t),
\end{equation}
where $\Gamma$ is a constant phenomenological relaxation matrix that accounts for the interaction with the bath by population relaxation
($\Gamma_{nn} = \frac{1}{\tau_1}$, where $\tau_1 = 100$ fs)
and coherence dephasing ($\Gamma_{nm} = \frac{1}{\tau_2}$, where $\tau_2=2$ ps). We tested longer population relaxation times up to 10 ns, and we found no change in the computed TTR response.

The time-varying Hamiltonian, $H(t)$, describes interactions of the system with the THz fields,
\begin{equation}
H(t;t_1) = H - M\cdot E_T(t) - M\cdot E_T(t-t_1),
\end{equation}
where $E_T(t)$ is the experimentally measured pulse shape \cite{finneran20172d}, $H$ is the system Hamiltonian, and $M$ is the transition dipole matrix.
Using $E_T(t)$ in the simulation, instead of a simple delta function $E\delta(t)$ ensures that the computed spectrum is in fact the molecular response convoluted with the instrument response function (IRF) and can be directly compared to the raw experimental data.

We propagate equation \ref{equLvN} forward in time by numerical integration, noting that the following mixed central-forward scheme provides good numerical stability:
\begin{equation}
\begin{aligned}
\rho(t+dt;t_1) &= \rho(t-dt;t_1)-\frac{2i}{\hbar}[H(t;t_1),\rho(t;t_1)]dt\\
               &-2\Gamma\rho(t-dt;t_1)dt,
\end{aligned}
\end{equation}
where the commutator is discretized by central difference, which preserves time reversibility, and the phenomenological dissipation is discretized by forward difference.
We then calculate the  TTR signal as the field emitted by the final Raman process,
\begin{equation}
S^{Th}(t_2;t_1) = \Re\left(\Tr(\Pi\cdot\rho(t_2;t_1)P)\right),
\end{equation}
where $\Pi$ is the transition polarizability, another important model parameter, and $P$ is a matrix with elements
\begin{equation}
P_{nm} = i, ~P_{mn} = -i, ~\text{and}~ P_{nn} = 1.
\end{equation}
$P$ ensures that the signal is emitted $90^\circ$ out-of-phase with respect to the induced polarization \cite{hamm2011concepts}.

We compute the frequency response from the absolute value of the Fourier transform, normalized  to its maximum value,
\begin{equation}
\tilde{S}^{Th}(f_2;f_1) = \lvert\FT(S^{Th}(t_2;t_1)\rvert
\end{equation}
For a given realization of parameters in $H$, $M$, and $P$,
comparison of the simulated and experimental TTR signals is made via the logarithmic error function:
\begin{equation}
\delta = \iint df^2 \left[log(\tilde{S}^{Ex}(f_2;f_1))-log(\tilde{S}^{Th}(f_2;f_1))\right]^2
\end{equation}
We fit the RDM model by minimizing $\delta$ with respect to $H$, $M$, and $P$, subject to $L_2$-norm regularization.  Specifically, we  minimize
\begin{equation}\label{TotErr}
\Delta = \delta + \lambda (h_i^2+\mu_j^2+\alpha_k^2),
\end{equation}
where $h_i$, $\mu_i$ and $\alpha_i$ are the parameters of the Hamiltonian $H=H(h_i)$, transition dipole $M=M(\mu_i)$ and transition polarizability $\Pi=\Pi(\alpha_i)$ operators.

We note that Eq.~\ref{equLvN} only accounts for Markovian dissipative processes.  This description may of course be extended to include more complex relaxation phenomena such as coherence and population transfer in the context of non-Markovian dynamics,\cite{steffen2000two,tanimura2000two,ishizaki2006modeling} which have been shown to be important in some liquids.

\subsection{Interpreting the Experimental Data}\label{secIIa}

The experimental setup\cite{finneran2016coherent} consists of two THz pump pulses, THz$_1$ and THz$_2$, which are polarized along the Y and X directions and separated by a time delay, $t_1=t_{\text{THz}_1}-t_{\text{THz}_2}$. A near-infrared (NIR) optical probe, polarized along X, induces a Y-polarized Raman response from the system at time delay $t_2=t_{\text{NIR}}-t_{\text{THz}_1}$. Due to the symmetry of the third-order response, a Raman signal is only detected after the second THz pump has interacted with the system. Experimentally, $t_1$ is scanned by changing the path length that the THz$_2$ field travels, while the THz$_1$ field remains fixed in time. A consequence of this design is that the third-order response will occur at $t_2=0$ when THz$_1$ is the second field interaction ($t_1>0$), but at a time $t_2=t_1$ when THz$_2$ is the second interaction ($t_1<0$). Qualitatively, a diagonally skewed TTR response is observed in the $t_1<0$ region. This must be properly accounted for in the interpretation of the spectrum. 

\begin{figure}[ht!]
\includegraphics[width=0.48\textwidth]{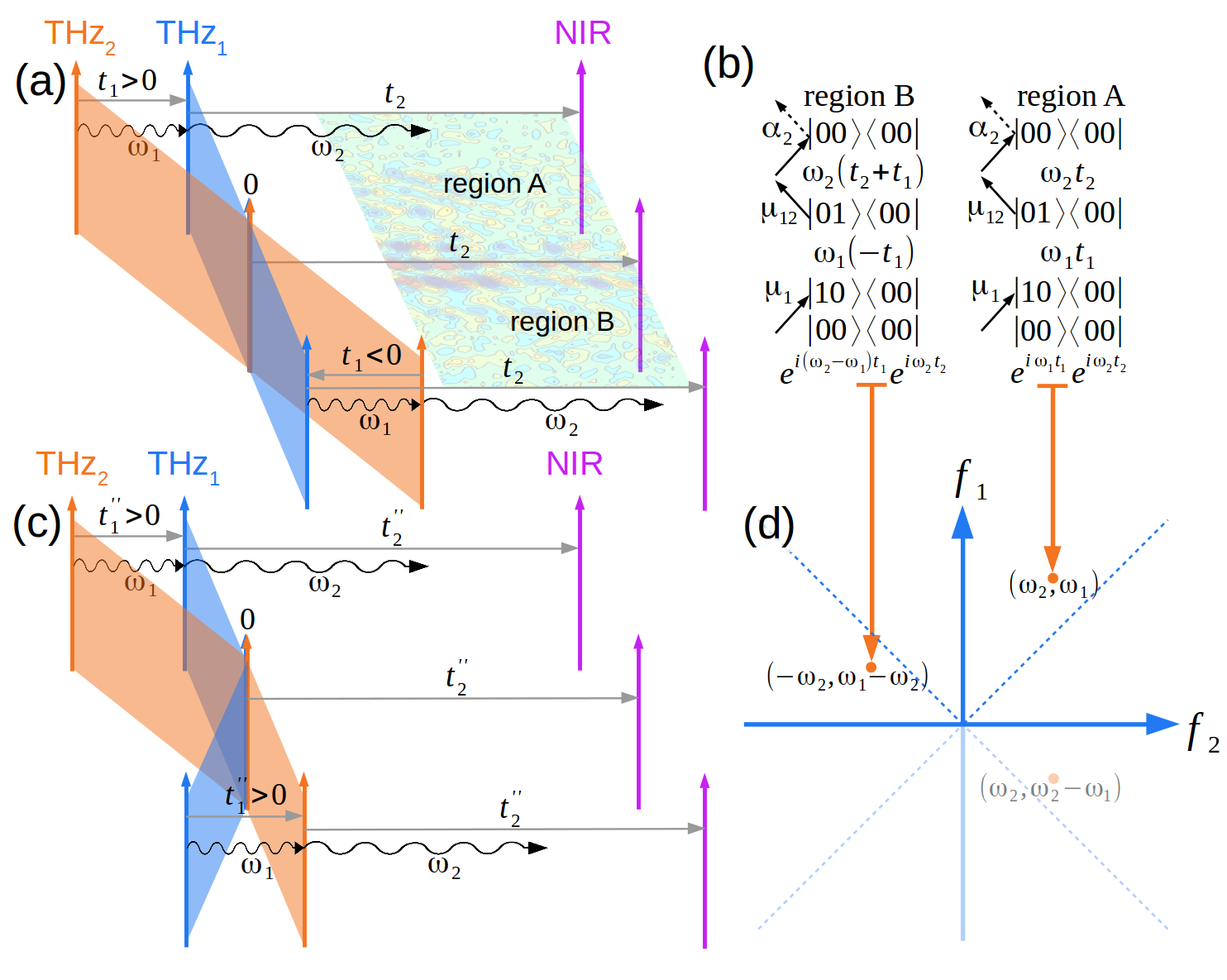}
\caption{\label{figMetA} Illustration of the THz pump pulses switching time ordering and the implication on the spectral features. Panel (a) shows schematically the experimental setup and the definition of time coordinates $t_1$ and $t_2$ in the two regions of response, A and B, corresponding to the pulse sequence THz$_2$-THz$_1$-NIR and THz$_1$-THz$_2$-NIR, respectively. Panel (b) illustrates a generic Liouville pathway represented by two distinct Feynman diagrams that feature different oscillatory phases in the portions of response A and B, as a consequence of mixing the time coordinates. These two Feynman diagrams which represent the same physical process manifest as two distinct peaks in the Fourier representation of the response: panel (d). Finally, panel (c) illustrates the transformed time coordinates $t_1^{''}$ and $t_2^{''}$ which fix this issue.}
\end{figure}

To illustrate the importance of correctly including this skew in the analysis,  we consider how a Liouville pathway in a generic system composed of two vibrational modes, changes when evaluated at $t_{1}>0$ and $t_{1}<0$. In Fig.~\ref{figMetA}(a), we label these two regions A and B, corresponding to pulse ordering THz$_2$-THz$_1$-NIR ($t_1>0$) and THz$_1$-THz$_2$-NIR ($t_1<0$), respectively. When the pathway is sampled in  region A of the response, the first THz pulse excites coherence $\lvert10\rangle\langle00\rvert$, which oscillates at  frequency $\omega_1$ and acquires  phase $e^{i\omega_1t_1}$, while the second THz pulse switches the coherence to $\lvert01\rangle\langle00\rvert$ which oscillates at frequency $\omega_2$ and acquires  phase $e^{i\omega_2t_2}$. Overall, this physical process generates a signal that is proportional to $e^{i\omega_1t_1}e^{i\omega_2t_2}$. The same pathway, when sampled in region B, generates a signal that is proportional to $e^{-i\omega_1t_1}e^{i\omega_2(t_2+t_1)}$, or rearranging, $e^{i(\omega_2-\omega_1)t_1}e^{i\omega_2t_2}$. Hence, taking the Fourier transform of the full $t_1$ domain (including both regions A and B)  results in two distinct peaks at $(\omega_2,\omega_1)$ and $(\omega_2,\omega_2-\omega_1)=(-\omega_2,\omega_1-\omega_2)$ that otherwise represent the same physical process. This is  undesirable and creates the opportunity for misinterpreting the TTR spectrum.

To ensure that each physical process is represented by a single distinct peak, we perform a coordinate transformation of the time response before computing the frequency response by FT. This transformation consists of skewing the response along $t_2$,
\begin{equation}
\label{equMetA1}
(t_2',t_1') = 
\begin{cases}
(t_2,t_1),&\text{region A} \\
(t_2+t_1,t_1),&\text{region B}
\end{cases}
\end{equation}
and then flipping it with respect to $t_1$,
\begin{equation}
\label{equMetA2}
(t_2'',t_1'') = 
\begin{cases}
(t_2',t_1')=(t_2,t_1),&\text{region A} \\
(t_2',-t_1')=(t_2+t_1,-t_1),&\text{region B}
\end{cases}
\end{equation}
as shown in Fig.~\ref{figMetA}(c). With these transformed time coordinates, our illustrative Liouville pathway  acquires the same phase factors $e^{i\omega_1t_1''}e^{i\omega_2t_2''}$ when sampled in region B as in region A. This transformation simplifies the resulting TTR spectrum and eliminates the appearance of redundant peaks.

\subsection{Enforcing Inversion Symmetry in the RDM Model}\label{secIIb}

Liquids are isotopic, such that any response function,
\begin{equation}
S(E) = \varepsilon_1 E+\varepsilon_2 E^2+\varepsilon_3 E^3+\varepsilon_4 E^4+\ldots,
\end{equation}
must obey inversion symmetry with respect to an applied electric field, $E$. The TTR response must also obey this symmetry,
\begin{equation}
S(-E_T,-E_T,-E_{NIR}) = -S(E_T,E_T,E_{NIR}),
\end{equation}
and requires that even-order contributions to the experimentally observed response must vanish,
\begin{equation}
\varepsilon_2 = \varepsilon_4 = \ldots = 0.
\end{equation}
This means that only scattering processes that involve an odd number of photons, not counting the final emission, can generate an experimental TTR signal. In the case of liquid bromoform, the final Raman process is linear and it only involves a one-photon absorption \cite{finneran20172d}. Therefore, the total number of THz photon interactions must be even and the leading contribution to the TTR signal is 3$^{rd}$ in the field, involving the absorption of two THz photons and one NIR photon. In other words, the TTR response scales linearly with each of the THz fields.

\begin{figure*}[ht!]
\includegraphics[width=0.99\textwidth]{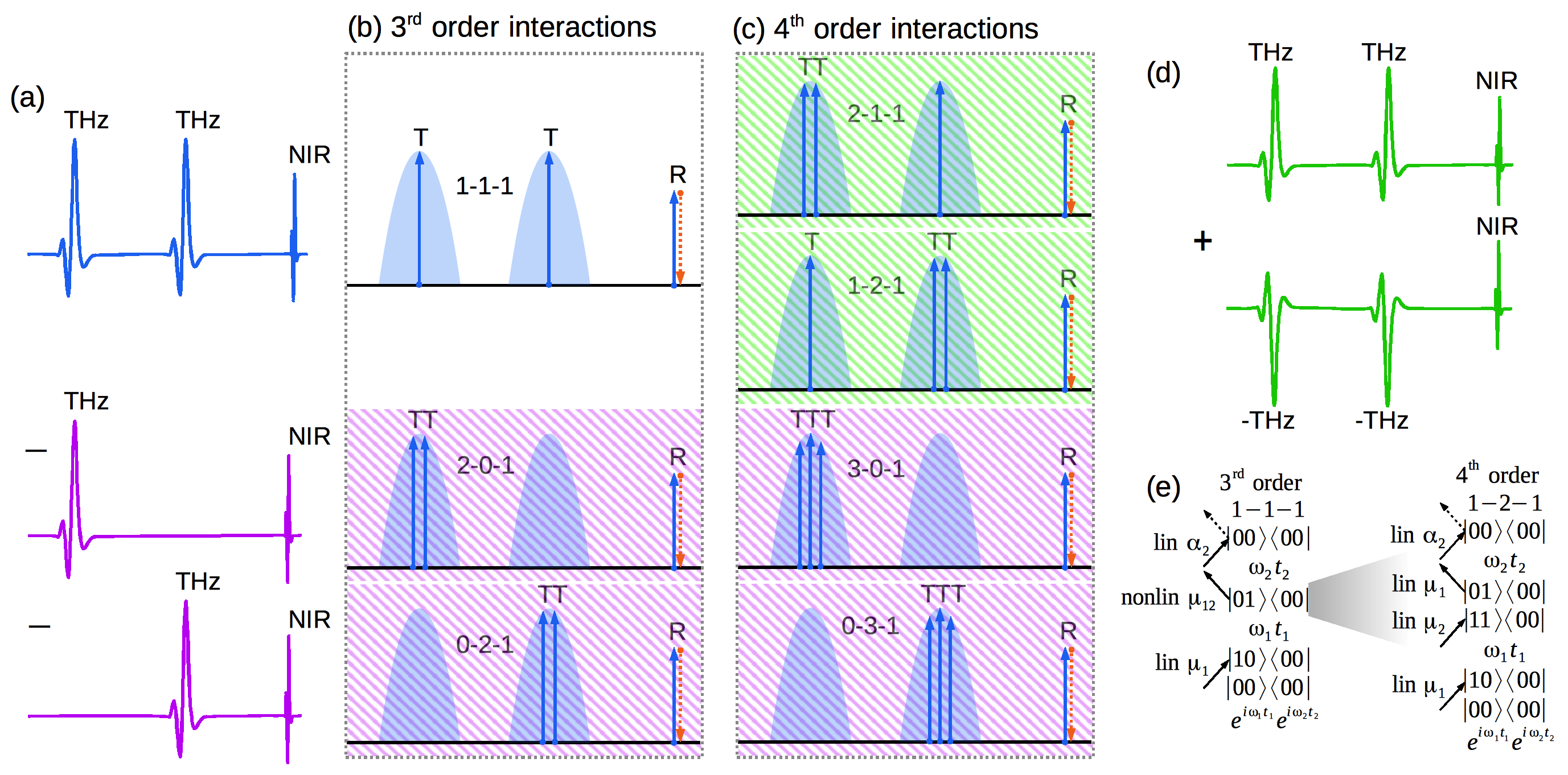}
\caption{\label{figMetB} Differential chopping and enforced inversion symmetry remove unwanted responses from the spectrum generated via RDM. Panel (a) schematically illustrates differential chopping, while panel (d) shows how we enforce inversion symmetry. Panels (b) and (c) list all 3$^{rd}$ and 4$^{th}$ order field interactions manifested in the RDM, from which only the 1-1-1 process is sampled experimentally. Shaded in purple are the processes removed by differential chopping, and in green the processes removed by enforcing  inversion symmetry. Panel (e) exemplifies two similar Liouville pathways of 3$^{rd}$ and 4$^{th}$ order, in which the transfer of coherence from $\lvert10\rangle \langle00\rvert$ to $\lvert01\rangle \langle00\rvert$ takes place via one nonlinear dipole interaction in the former and two linear dipole interactions in the latter.}
\end{figure*}

Fig.~\ref{figMetB}(b) lists all possible 3$^{rd}$ order interactions in which two THz photons are involved: 1-1-1, 2-0-1 and 0-2-1. The TTR experiment further isolates the desired 1-1-1 response by employing differential chopping \cite{finneran2016coherent}. In the RDM simulations, we implement differential chopping by separating the 1D responses $S_2^{Th}$ and $S_1^{Th}$ from the total response $S^{Th}$, as illustrated in Fig.~\ref{figMetB}(a). We compute these individual responses by propagating the dynamics under the partial Hamiltonians
\begin{equation}
\begin{aligned}
&H_2(t;t_1) = H - M\cdot E_T(t)~~\text{and}\\
&H_1(t;t_1) = H - M\cdot E_T(t-t_1).
\end{aligned}
\end{equation}
This simulation procedure removes the unwanted field interactions, shaded purple in Fig.~\ref{figMetB}, from the 3$^{rd}$ response. However, the RDM simulated response \cite{finneran2016coherent,finneran20172d} also includes 4$^{th}$ order interactions, unlike the experiment where even order processes vanish due to the symmetry of the liquid, as explained above.

It is found that for the simulated TTR response, 4$^{th}$ order contributions can be comparable in magnitude to 3$^{rd}$ order ones, because the latter require at least one nonlinear dipole interaction\footnotemark[2] while the former do not. For instance, in Fig.~\ref{figMetB}(e), we compare two similar pathways of 3$^{rd}$ and 4$^{th}$ order. In the 3$^{rd}$ order process, the two-quanta excitation from $\lvert10\rangle$ to $\lvert01\rangle$ is achieved by a single photon interacting with a nonlinear dipole element and has a scattering amplitude proportional to
\begin{equation}
\sigma_{3^{rd}} \propto \langle 10 \rvert \frac{\partial^2 M} {\partial Q_1 \partial Q_2} \lvert01\rangle E_{T}.
\end{equation}
Meanwhile, in the 4$^{th}$ order process, the same excitation is achieved by two consecutive photons acting on the linear dipole via an intermediate state,
\begin{equation}
\sigma_{4^{th}} \propto \langle 10 \rvert \frac{\partial M} {\partial Q_2} \lvert11\rangle \langle 11 \rvert \frac{\partial M} {\partial Q_1} \lvert01\rangle E^2_{T}.
\end{equation}
For large THz fields, this ``all-linear" 4$^{th}$ order can become larger than the 3$^{rd}$ order if the system has small dipole nonlinearities, i.e.,
\begin{equation}
\frac{\partial^2 M} {\partial Q_1 \partial Q_2} \lesssim E_{T} \frac{\partial M} {\partial Q_2}  \frac{\partial M} {\partial Q_1}.
\end{equation}

\footnotetext[2]{To prove that all 3$^{rd}$ order TTR pathways require a nonlinear dipole interaction, consider a 3$^{rd}$ order Liouville pathway which starts from a generic population state $\lvert a \rangle \langle a \rvert$. Interaction of the system  with the first THz pulse  changes the excitation state of the bra or ket of the density matrix by $n$ quanta to a  $\lvert a+n \rangle \langle a \rvert$ coherence. Then, upon a second THz interaction, the system changes to either $\lvert a+n+m \rangle \langle a \rvert$ or $\lvert a+n \rangle \langle a+m \rvert$. Since the final Raman interaction only changes the system by a single quantum \cite{finneran20172d}, and since all pathways must end in a population state, it follows that either $\lvert n+m \rvert=1$ or $\lvert n-m \rvert=1$. All paths involving $n=0$ or $m=0$ are removed by differential chopping, and therefore all surviving diagrams have both nonzero $n$ and $m$. As a result, at least one of the THz processes exchanges more than one quantum with the systems, and it is strictly nonlinear.}

To address this issue, 4$^{th}$ order contributions to the simulated RDM response must be explicitly removed. We achieve this by simulating both the response to positive THz fields and negative THz fields, and then summing the two contributions as illustrated in Fig.~\ref{figMetB}(d),
\begin{equation}
S^{Th} = S^{Th}(E_{T},E_{T}) + S^{Th}(-E_{T},-E_{T}).
\end{equation}
This procedure  ensures that the final response function is odd with respect to the overall field and that the 4$^{th}$ order contributions, shaded green in Fig.~\ref{figMetB}(c), vanish. As a result, the simulated TTR spectrum correctly reflects the inversion symmetry of the liquid state and can be compared directly to the experimental spectrum.

\subsection{The Hamiltonian, Dipole, and Polarizability Surfaces}\label{secIIc}

\begin{figure*}[ht!]
\includegraphics[width=0.99\textwidth]{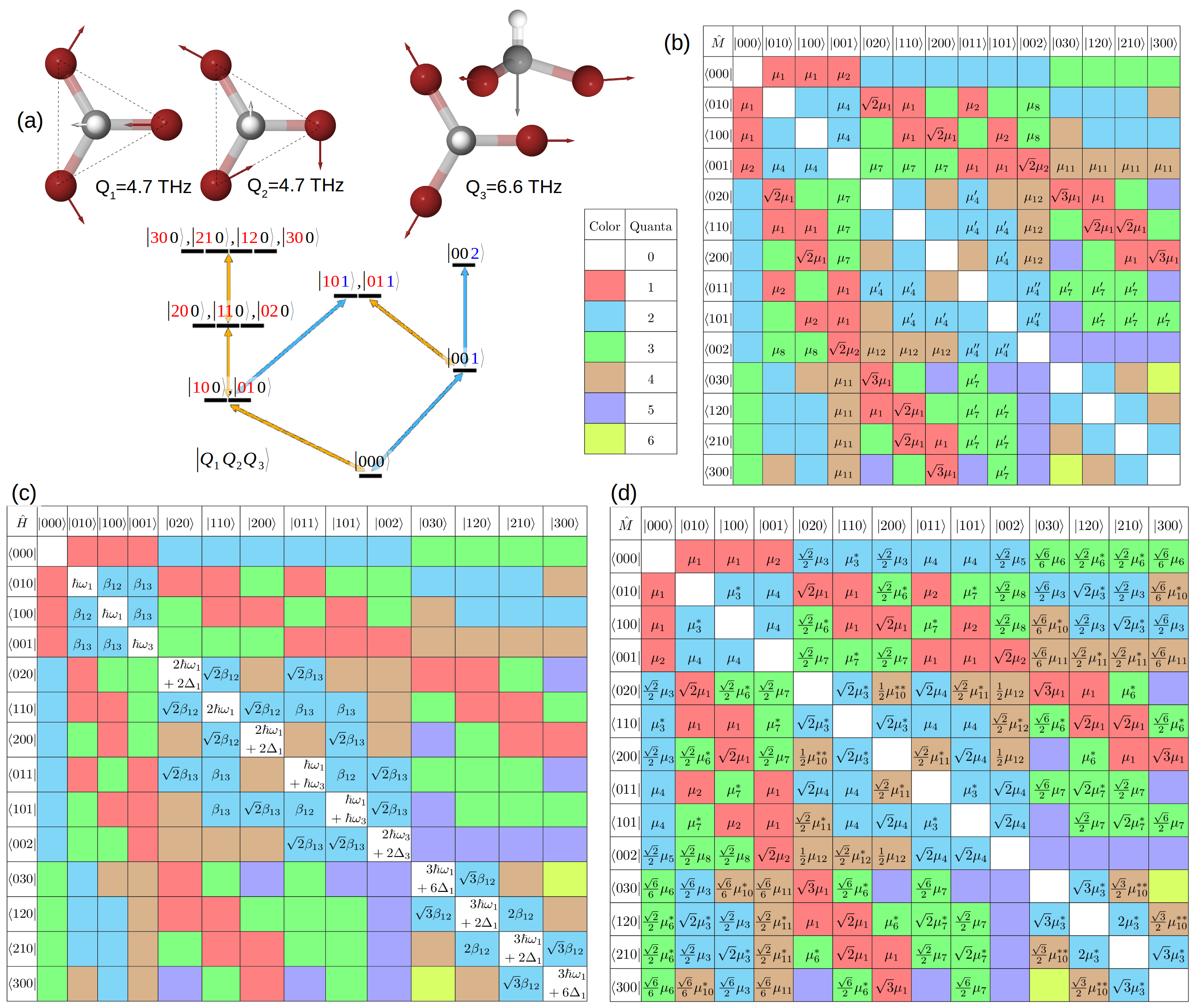}
\caption{\label{figMetC}
Panel (a) shows the low energy normal modes of the bromoform molecule: the degenerate ``C-Br bending modes'' $Q_1$ and $Q_2$ at 4.7 THz and the ``umbrella mode'' $Q_3$ at 6.6 THz. The panel also shows the corresponding quantum states labeled $\lvert Q_1 Q_2 Q_3 \rangle$ that we use to describe these normal modes. The double arrow indicates all possible 1-quantum transitions between these states. Panel (b) shows the parametrization of the transition dipole moment ($\hat{M}$) used in Ref. \citenum{finneran20172d}. The color assigned to each matrix element corresponds to the number of quanta involved in that transition as summarized in the inset table. Elements that appear empty are fixed to zero. Panels (c) and (d) illustrate the Hamiltonian ($\hat{H}$) and transition dipole ($\hat{M}$) in the current RDM model, as derived by integrating equations \ref{equHLO} and \ref{equMLO}. Notice that in this work the dipole is fully parameterized to the 4$^{th}$ order and only transition matrix elements higher than 4$^{th}$ order are set to zero.}
\end{figure*}

The bromoform molecule has three vibrational degrees of freedom in the terahertz regime: two degenerate C-Br bending modes at 4.7 THz, $Q_1$ and $Q_2$, and a symmetric umbrella mode at 6.6 THz, $Q_3$ \cite{fernandezliencres1996the}.
Fig.~\ref{figMetC}(a) illustrates these motions and provides an energy-level diagram with the number of quanta in each mode indicated, using notation $\lvert Q_1 Q_2 Q_3 \rangle$. In typical 2D-IR spectroscopy applications at room temperature, only the ground and first-excited state of each mode are involved. TTR spectroscopy probes modes that are low in energy, such that even at room temperature ($kT\approx6.2$ THz), it is necessary to consider thermally accessible excited states. In the current study, the bandwidth of the THz pulse covers  $BW\approx8$ THz of energy, such that all the states below $kT+BW\lesssim14.2$ THz should be considered. The highest energy state that is included in our model is the quadruplet $Q_1\otimes Q_2$, amounting to three quanta of energy in the  degenerate mode ($3\times 4.7$ THz = $14.1$ THz).

The calculated TTR spectrum depends on the parameterization of $H=H(h_i)$, $M=M(\mu_i)$ and $\Pi=\Pi(\alpha_i)$. In the RDM model of Ref. \citenum{finneran20172d}, we employed a harmonic Hamiltonian,
\begin{equation}
\begin{aligned}
\label{equHWF}
\hat{H} &= \hbar \omega_1 \left(\hat{a}_1^\dagger \hat{a}_1+\hat{a}_2^\dagger \hat{a}_2\right) + \hbar \omega_3 \hat{a}_3^\dagger \hat{a}_3,
\end{aligned}
\end{equation}
the parameters for which were fixed to the linear absorption experimental spectrum, and a transition dipole of the form
\begin{equation}
\begin{aligned}
\label{equMWF}
\hat{M} &= \mu_1 \left(\hat{Q}_1+\hat{Q}_2\right) + \mu_2 \hat{Q}_3 + \hat{M}_{2q}(\mu_4,\mu_4',\mu_4'') \\
 &+ \hat{M}_{3q}(\mu_7,\mu_7',\mu_8) + \hat{M}_{4q}(\mu_{11},\mu_{12}),
\end{aligned}
\end{equation}
where $\hat{M}_{nq}$ represent the nonlinear blocks of the dipole operator comprised of $n$-quanta $(n>1)$ transitions, which were determined by fitting to the nonlinear TTR experimental spectrum. The $\hat{Q}_i$ are non-dimensional normal modes which can be expressed in terms of the creation and annihilation operators $\hat{a}_i^\dagger$ and $\hat{a}_i$,
\begin{equation}
\begin{aligned}
\label{equQ1}
\hat{Q}_i = \sqrt{\frac{2 m_i \omega_i}{\hbar}} \hat{q}_i = \hat{a}_i^\dagger+\hat{a}_i.
\end{aligned}
\end{equation}
Combining Eqs.~\ref{equQ1} and \ref{equMWF} and integrating the linear part of the dipole in the $\lvert Q_1 Q_2 Q_2 \rangle$ basis leads to the matrix parameterization of Ref.~\citenum{finneran20172d} (Fig.~\ref{figMetC}(b)).

In this work, we develop a more comprehensive model for the Hamiltonian, which includes the anharmonicity $\Delta_i$ of the single molecule normal modes, as well as mode coupling $\beta_{ij}$ originating from the condensed-phase environment. We employ a quantum-conserving Hamiltonian of the form
\begin{strip}
\begin{equation}
\begin{aligned}
\label{equHLO}
\hat{H} &= \hbar \omega_1 \left(\hat{a}_1^\dagger \hat{a}_1+\hat{a}_2^\dagger \hat{a}_2\right) + \hbar \omega_3 \hat{a}_3^\dagger \hat{a}_3 + \beta_{12} \left(\hat{a}_1^\dagger \hat{a}_2 + \hat{a}_1 \hat{a}_2^\dagger \right)+\beta_{13} \left(\hat{a}_2^\dagger \hat{a}_3 + \hat{a}_2 \hat{a}_3^\dagger + \hat{a}_1^\dagger \hat{a}_3 + \hat{a}_1 \hat{a}_3^\dagger \right) \\
&{}+\Delta_1 \left(\hat{a}_1^\dagger{}^2 \hat{a}_1^2+\hat{a}_2^\dagger{}^2 \hat{a}_2^2 \right) + \Delta_3 \hat{a}_3^\dagger{}^2 \hat{a}_3^2,
\end{aligned}
\end{equation}
\end{strip}
and integrate it in the $\lvert Q_1 Q_2 Q_2 \rangle$ basis to obtain the matrix representation shown in Fig.~\ref{figMetC}(c). Note that the couplings between the doubly degenerate modes $Q_{1,2}$ and mode $Q_3$ are equivalent and labeled $\beta_{13}$ in Eq.~\ref{equHLO}.

Furthermore, we Taylor expand the dipole operator to 4$^{th}$ order in the normal modes and enforce the $Q_{1,2}$ symmetry by grouping together terms which are invariant to label exchange (1$\leftrightarrow$2):
\begin{strip}
\begin{equation}
\begin{aligned}
\label{equMLO}
\hat{M} &= \mu_1 \left(\hat{Q}_1+\hat{Q}_2\right) + \mu_2 \hat{Q}_3 + \frac{1}{2!} \mu_3 \left(\hat{Q}_1^2+\hat{Q}_2^2\right) + \mu_3^{*} \hat{Q}_1 \hat{Q}_2 + \mu_4 \left(\hat{Q}_1+\hat{Q}_2\right)\hat{Q}_3 + \frac{1}{2!} \mu_5 \hat{Q}_3^2\\
&{}+ \frac{1}{3!}\mu_6 \left( \hat{Q}_1^3+\hat{Q}_2^3\right) + \frac{1}{2!} \mu_6^{*}  \hat{Q}_1 \hat{Q}_2 \left(\hat{Q}_1+\hat{Q}_2\right) + \frac{1}{2!} \mu_7 \left(\hat{Q}_1^2+\hat{Q}_2^2\right)\hat{Q}_3 + \mu_7^{*}  \hat{Q}_1 \hat{Q}_2 \hat{Q}_3 + \frac{1}{2!} \mu_8 \left(\hat{Q}_1+\hat{Q}_2\right)\hat{Q}_3^2\\
&{}+ \frac{1}{3!} \mu_{10}^{*}  \hat{Q}_1 \hat{Q}_2 \left(\hat{Q}_1^2+\hat{Q}_2^2\right) + \frac{1}{2!2!} \mu_{10}^{**}\hat{Q}_1^2 \hat{Q}_2^2 + \frac{1}{3!} \mu_{11} \left(\hat{Q}_1^3+\hat{Q}_2^3\right)\hat{Q}_3 + \frac{1}{2!} \mu_{11}^{*}  \hat{Q}_1 \hat{Q}_2 \left(\hat{Q}_1+\hat{Q}_2\right) \hat{Q}_3 \\
&{}+ \frac{1}{2!2!} \mu_{12} \left(\hat{Q}_1^2+\hat{Q}_2^2\right)\hat{Q}_3^2 + \frac{1}{2!} \mu_{12}^{*}  \hat{Q}_1 \hat{Q}_2 \hat{Q}_3^2.
\end{aligned}
\end{equation}
\end{strip}
Whereas the dipole form in Ref.~\citenum{finneran20172d} (Eq.~\ref{equMWF}) uses a normal mode expansion of the linear part and an empirical parameterization of the nonlinear part, the current work expands the full dipole operator in the normal modes to 4$^{th}$ order. As a result, the parameters of the current model are associated with the partial derivatives of $\hat{M}$ with respect to the $\hat{Q}_i$. 
We express the $n^{th}$ power of $\hat{Q}_i$ as creation-annihilation $n$-tuples,
\begin{equation}
\begin{aligned}
\label{equQn}
\hat{Q}_i^n &= \hat{a}_i^\dagger{}^n + \sum_{perm} \hat{a}_i^\dagger{}^{(n-1)}\hat{a}_i + \sum_{perm} \hat{a}_i^\dagger{}^{(n-2)}\hat{a}_i^2 \\
&+\dotso+\sum_{perm}\hat{a}_i^\dagger{}^2\hat{a}_i^{(n-2)}+\sum_{perm}\hat{a}_i^\dagger{}\hat{a}_i^{(n-1)} + \hat{a}_i^n
\end{aligned}
\end{equation}
and we keep only the pure creation (first term) and pure annihilation (last term); all other terms account for transitions involving less than $n$ quanta.
We then integrate Eq.~\ref{equMLO} in the $\lvert Q_1 Q_2 Q_2 \rangle$ basis and obtain the model illustrated in Fig.~\ref{figMetC}(d). The polarizability operator is parameterized in a similar way; however, it is only expanded to  second power in $Q_i$, since the final Raman interaction is expected to be linear.\cite{finneran20172d}

To fit the simulated RDM spectrum to the experimental TTR result, we keep $\hbar\omega_1$, $\hbar\omega_3$, $\mu_1$, $\mu_3$, and $\alpha_1$ fixed, while varying the remaining parameters. We perform the RDM dynamics in the diagonal representation of the Hamiltonian. As such, we diagonalize $H$,
\begin{equation}
H'=D^\dagger{} H D
\end{equation}
and apply the same transformation to the dipole and polarizability,
\begin{equation}
M'=D^\dagger{} M D,\:\:\Pi'=D^\dagger{} \Pi D.
\end{equation}
performing the RDM simulations with $H',M'$ and $P'$.

\section{Results and Discussions}

The results are presented in two parts. We first present the revised experimental spectrum obtained through the transformation of the time coordinates.We then discuss the results of the RDM model introduced in this work.

\subsection{Revised Experimental Spectrum}

The raw experimental TTR response is comprised of an orientational molecular response with superimposed oscillations arising from vibrational coherences. As in previous work\cite{allodi2015nonlinear, finneran2016coherent}, we focus on the vibrational component of the response, so we use an exponential fit along $t_{2}$ to de-trend the orientational response from the data.

\begin{figure*}[ht!]
\includegraphics[width=0.99\textwidth]{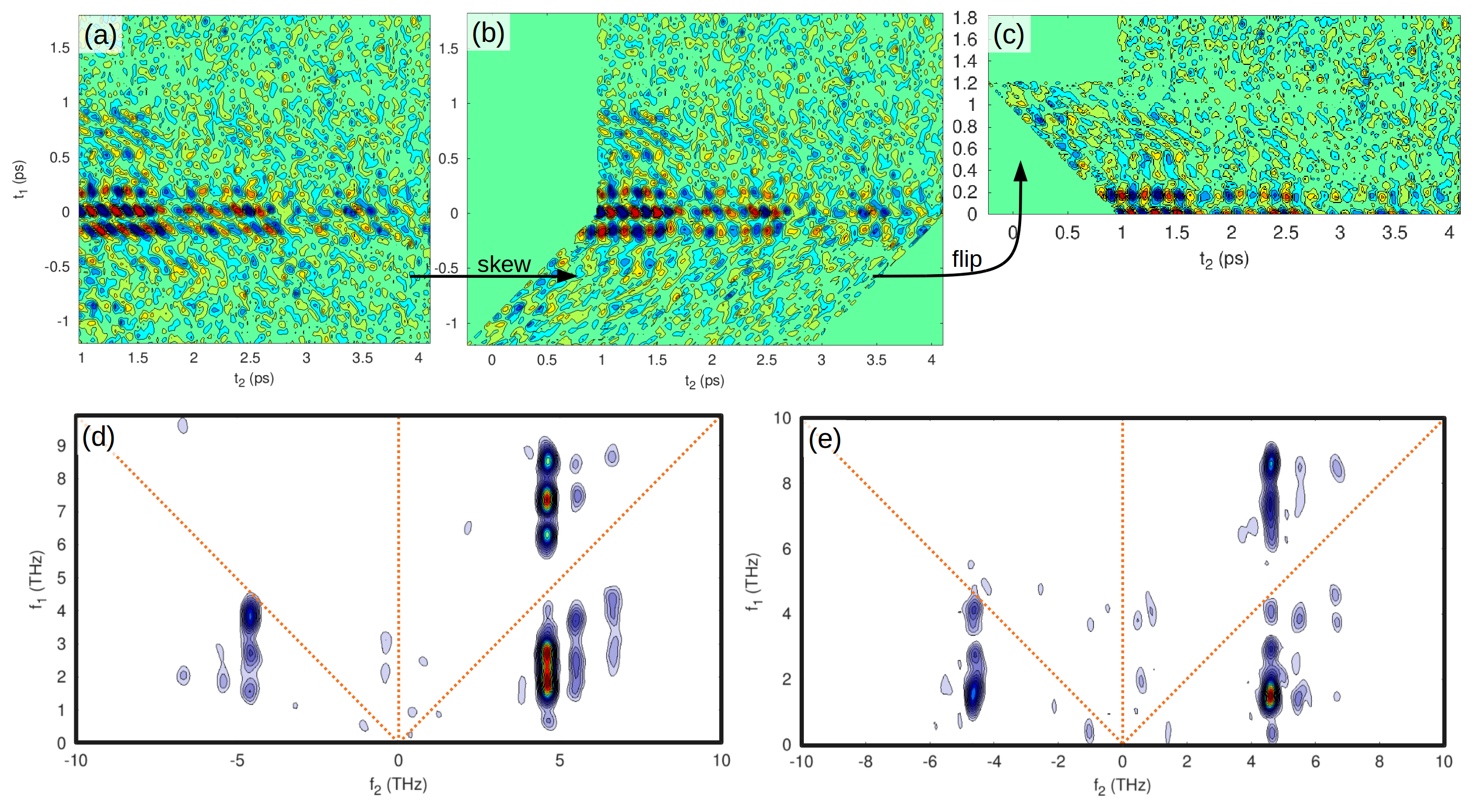}
\caption{\label{figResA} Illustration of the effect of the introduced time-coordinate  transformation on the experimental spectrum. Panels (a), (b) and (c) show how the time transformation (skew and flip) alters the response in time domain. Panel (a) corresponds to the initial coordinates represented schematically in Fig.~\ref{figMetA}(a), while panel (c) to the proposed coordinates represented in Fig.~\ref{figMetA}(c). Panels (d) and (e) present the Fourier transforms of panels (a) and (c), before and after the time transformation, respectively.}
\end{figure*}

In Figs.~\ref{figResA}(a)-(c), we demonstrate how the vibrational signal changes upon application of the time-coordinate transformation from section Interpreting the Experimental Data (\ref{secIIa}). The first step of the transformation involves a  skew along $t_2$, which makes the response function appear symmetric with respect to the $t_1=0$ line. The transformed signal is symmetric because swapping the order of the THz pulses does not change the response type; in both cases, we measure a TTR signal. This is unlike a TRT experiment where inverting the pulse order results in a RTT sequence which measures a different correlation function altogether \cite{savolainen2013two-dimensional}.
The second time transformation involves a flip with respect to $t_1$, which due to the symmetry of the response about $t_1=0$ simply averages the A and B regions of the data and improves the signal to noise ratio.

Panels (d) and (e) of Fig.~\ref{figResA} show the absolute value Fourier transforms of the original and  transformed time responses, respectively. The revised spectrum is significantly simplified (panel (e)). Whereas the original  spectrum features six intense peaks, the revised spectrum shows three features that are familiar in the context of conventional two dimensional spectroscopy. The most intense peak forms at $f_1=\omega_3-\omega_1$ and reports on the coupling between modes $Q_1$ and $Q_3$ via a nonlinear dipole interaction. The equivalent mode in the negative $f_2<0$ quadrant is closely related and shows a clear rephasing pathway. Note that in the revised spectrum, the $f_2<0$ quadrant corresponds to true rephasing, whereas in the original spectrum the peaks involve a mix of rephasing and higher overtone non-rephasing pathways. Interestingly, the most intense true-rephasing feature appears tilted at an angle; this tilt could be related to the degree of inhomogeneous broadening. 
Finally, in the non-rephasing quadrant along the first diagonal, the higher frequency features appear symmetric to those at lower frequency. 

\subsection{Parameterization of the RDM Model}

We fit the parameters of the RDM model with respect to the revised experimental spectrum using the Levenberg-Marquardt algorithm implemented in Octave \cite{levenberg1944method,marquardt1963algorithm,octave}, minimizing the penalty function in equation \ref{TotErr}. For each value of the regularization parameter $\lambda$, we start with random initial guesses for the RDM parameters, sampled  uniformly from the $\left[-1,1\right]$ interval. The fitting problem is highly under-determined, featuring multiple local minima, due to the large number (order $10^4$) of Liouville paths that compete in the creation of a relatively small number of peaks. Additionally, the dipole and polarizability parameters can take negative values, which lead to peak cancellations. To address these challenges, we perform multiple minimizations, starting each time from a different random initial guess.

\begin{figure*}[ht!]
\includegraphics[width=0.99\textwidth]{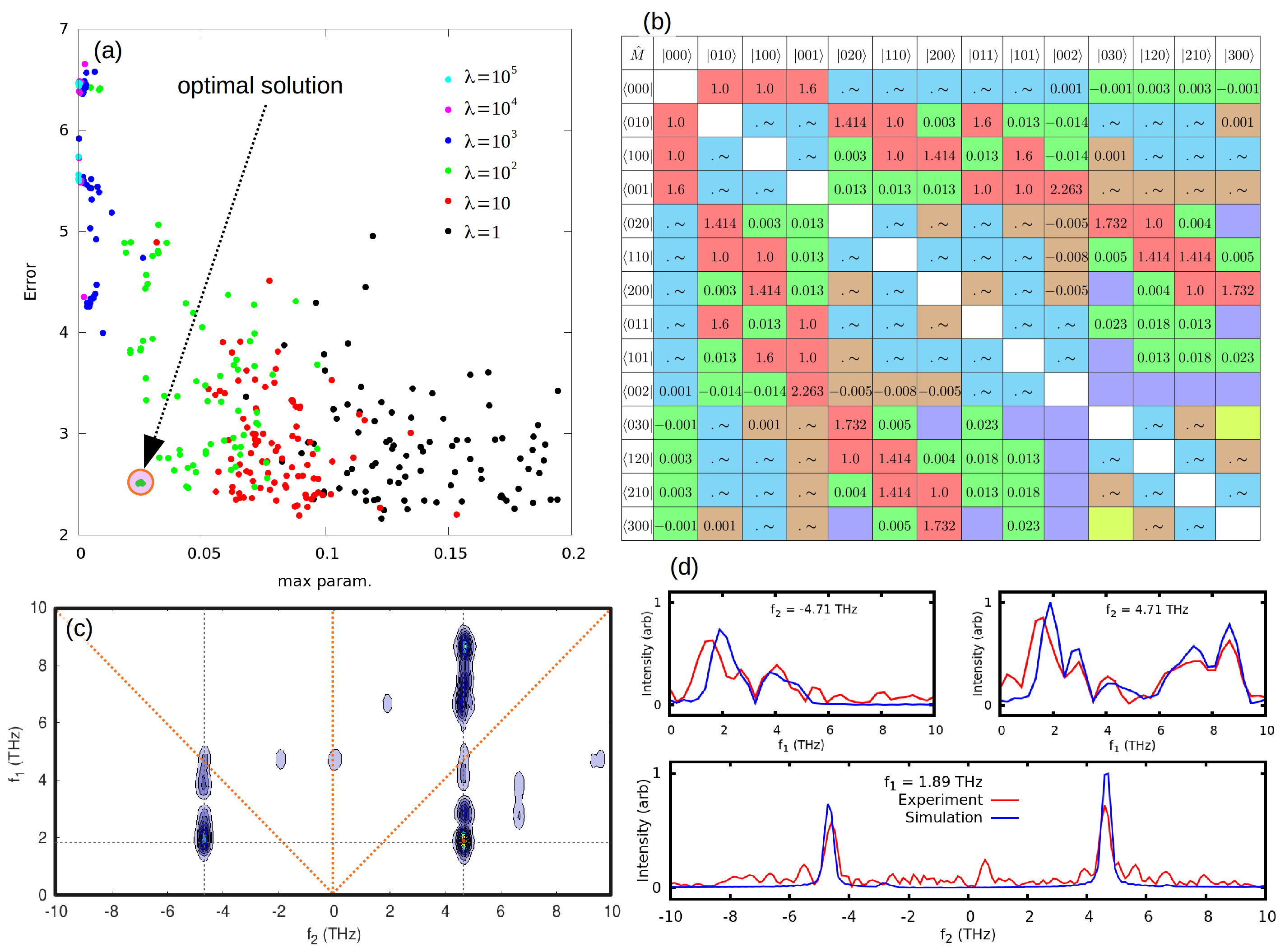}
\caption{\label{figResB} Fitting the parameters of the RDM model to experiment. Panel (a) illustrates the regularization procedure. Each point is a local solution represented by a set of fitted parameters ${p}_i$. Here we plot the error of each solution $\Delta E_i$ as function of the maximum parameter value, $max({p}_i)$. The color represents the different regularization coefficient $\lambda$. The optimal solution minimizes the error with the smallest possible parameters. Panels (b) and (c) show the transition dipole matrix, the simulated spectrum corresponding to the optimal solution identified in panel (a). In panel (b) we use the placeholder ``$.\thicksim$'' for parameters that are essentially zero within three significant digits. Panel (d) shows three important 1D slices through the 2D spectrum at $f_1=1.89$ THz and $f_2=+-4.71$ THz in order to facilitate the comparison to the experimental result. The fit was performed with respect to the revised experimental spectrum shown in Fig.~\ref{figResA}(d).}
\end{figure*}

In Fig.~\ref{figResB}(a) we plot all the converged solutions as function of the maximum fitted parameter $max(h_i,\mu_j,\alpha_k)$, and color them according to the $L_2$ regularization strength, $\lambda$. Clustered to the left are under-fitted results with small fitting parameters but large errors, while to the right are the over-fitted results that match the spectrum at the expense of unphysical parameters. The optimal solution lies in the intermediate regime with $\lambda=100$, as indicated in Fig.~\ref{figResB}(a). Three independent minimizations starting from different initial guesses converged to the same error with very similar parameters, suggesting relative robustness of this optimal paremeterization. We summarize the best-fit parameters in tables \ref{tab01} and \ref{tab02}, which report the values of the transition dipole elements, and we show the full transition dipole matrix in Fig.~\ref{figResB}(b).

\begin{table*}
\caption{\label{tab01} Values of the transition dipole matrix, obtained from \emph{ab initio} (CCSD/aug-cc-pVTZ) calculations\cite{finneran20172d} and via fitting of the RDM models in Ref.~\citenum{finneran20172d} and in the current work.}
\begin{tabular}{cc|c|l|l}
\hline
\hline
$ \langle i \rvert \hat{M} \lvert j \rangle$&\#Q&CCSD\cite{finneran20172d}& RDM (Ref. \citenum{finneran20172d})& RDM (current)\\
\hline
$\lvert 000 \rangle \rightarrow \lvert 100 \rangle$ & 1 & 1.00 & $\phantom{-}1.00$ (fixed) & $\phantom{-}1.00$ (fixed) \\
$\lvert 000 \rangle \rightarrow \lvert 001 \rangle$ & 1 & 1.01 & $\phantom{-}1.60$ (fixed) & $\phantom{-}1.60$ (fixed) \\
\hline
$\lvert 100 \rangle \rightarrow \lvert 001 \rangle$ & 2 & 0.04 & $\phantom{-}0.009\left(\mu_4\right)$ & $\phantom{-}0.000\left(\mu_4\right)$ \\
$\lvert 200 \rangle \rightarrow \lvert 101 \rangle$ & 2 & 0.06 & $\phantom{-}0.000\left(\mu_4'\right)$ & $\phantom{-}0.000\left(\sqrt{2}\mu_4\right)$ \\
$\lvert 101 \rangle \rightarrow \lvert 002 \rangle$ & 2 & 0.06 & $\phantom{-}0.000\left(\mu_4''\right)$ & $\phantom{-}0.000\left(\sqrt{2}\mu_4\right)$ \\
\hline
$\lvert 100 \rangle \rightarrow \lvert 002 \rangle$ & 3 & 0.00 & $\phantom{-}0.364\left(\mu_8\right)$ & $-0.014\left(\frac{\sqrt{2}}{2}\mu_8\right)$ \\
$\lvert 001 \rangle \rightarrow \lvert 200 \rangle$ & 3 & 0.01 & $\phantom{-}0.442\left(\mu_7\right)$ & $\phantom{-}0.013\left(\frac{\sqrt{2}}{2}\mu_7\right)$ \\
$\lvert 101 \rangle \rightarrow \lvert 300 \rangle$ & 3 & 0.02 & $\phantom{-}0.034\left(\mu_7'\right)$ & $\phantom{-}0.023\left(\frac{\sqrt{6}}{2}\mu_7\right)$ \\
\hline
$\lvert 001 \rangle \rightarrow \lvert 300 \rangle$ & 4 & - & $\phantom{-}0.023\left(\mu_{11}\right)$ &  $\phantom{-}0.000\left(\frac{\sqrt{6}}{6}\mu_{11}\right)$ \\
$\lvert 200 \rangle \rightarrow \lvert 002 \rangle$ & 4 & - & $\phantom{-}0.092\left(\mu_{12}\right)$ & $-0.005\left(\frac{1}{2}\mu_{12}\right)$ \\
\hline
\hline
\end{tabular}
\end{table*}

The resulting RDM simulated spectrum shown in Fig.~\ref{figResB}(c) is in excellent agreement with the revised experimental spectrum (Fig.~\ref{figResA}(e)). This can be clearly seen in Fig.~\ref{figResB}(d), which provides a side-by-side comparison between simulation and experiment along the three most  important 1D traces from the 2D response.

 It is found that use of the experimental pulse shape $E_T(t)$ in the RDM simulations is crucial for reproducing the experimental spectrum. Using simple $\delta$-functions to simulate the response, otherwise with the same RDM parameters, gives a radically different result, with just one dominant peak on the diagonal at $f_2=4.7$ THz. 

The dipole nonlinearities fitted with the RDM model in this work are significantly smaller than those found in Ref. \citenum{finneran20172d}, and they agree well with results from \textit{ab initio}  electronic structure calculations (see table \ref{tab01}). Critically, the third-order fitted non-linearities in the current RDM model have the same order of magnitude as those from the \emph{ab initio} electronic structure calculations, in contrast with the third-order terms from the fitted RDM model of Ref.~\citenum{finneran20172d}. This central result demonstrates that the refined treatment of the RDM model and the experimental spectrum in the current work removes the central inconsistency of Ref.~\citenum{finneran20172d}. 
Likewise, the fourth-order terms in the fitted RMD model of the current work are vastly smaller than those that were necessarily invoke in the previous study\cite{finneran20172d}.  For the second-order terms, both the current and earlier RDM models have very small fitted values. 
It is found that the difference between the fitted and \textit{ab initio} values for the second-order terms is insignificant. We confirmed this by running the current RDM model with the \textit{ab initio} values for $\mu_4$ and all other fitted parameters unchanged, and finding that the resulting simulated spectrum is unaffected.

\begin{table}
\caption{\label{tab02} Hamiltonian and transition polarizability parameters of the RDM model.}
\begin{tabular}{cc|cc}
\hline
\hline
$\hbar \omega_1$              & 4.7(fixed) & $\hbar \omega_3$              & 6.6(fixed) \\
$\Delta_{1} / \hbar \omega_1$ & -0.52\%    & $\Delta_{3} / \hbar \omega_3$ & -0.10\%    \\
$\beta_{12} / \hbar \omega_1$ & -0.05\%    & $\beta_{13} / \hbar \omega_3$ & +0.14\%    \\
\hline
$\alpha_1$                  & 1.00 (fixed) & $\alpha_2$                    & -0.003     \\
\hline
\hline
\end{tabular}
\end{table}

Finally, in Table \ref{tab02}, we report the fit RDM parameters for the anharmonicity $\Delta_i$ of the normal modes, their mechanical coupling $\beta_{ij}$, and the first order polarizability elements $\alpha_1$ and $\alpha_3$. We find small anharmonicities and mechanical coupling elements, which points to relatively harmonic, weakly coupled normal modes in bromoform. This conclusion is in agreement with 
vibrational second-order perturbation theory (VPT2) calculations performed previously.\cite{finneran20172d} 
Finally, the polarizability matrix elements obtained in this current work are in good agreement with those obtained in our previous work \cite{finneran20172d}, with very small nonlinear elements indicating that the final Raman interaction is indeed a linear process.

It is worth noting that the simulated spectrum has low-intensity peaks at $f_2 \simeq6.6$ THz that bear resemblance to those in the experimental spectrum. Given weakness of these peaks, they make a small contribution to the error function in the fitting and are thus poorly resolved. Additionally, in some fits, we found that the features at $f_2 \simeq5.5$ THz might be connected to the mechanical coupling between $Q_{1,2}$ and $Q_3$. Future refinement of the RDM model and higher resolution experiments are needed to interpret these subtle features.

\section{Conclusions}

Our original TTR study of bromoform \cite{finneran20172d} revealed large dipole nonlinearities which were inconsistent with \textit{ab initio} electronic structure calculations. Here, we refine both the experimental and theoretical description of this system, resolving the inconsistency. 
The key refinements are described below.

First, we revised the experimental spectrum by introducing a simplifying time-coordinate transformation. Due to the symmetry of the experimental setup, the time response must be skewed and flipped when the ordering of the two THz pulses is changed before taking the Fourier transform for a correct interpretation. The new spectrum reveals fewer features and true rephasing signals which are symmetric with their non-rephasing counterparts, as expected. 

Second, we have developed an RDM model that analytically includes all dipole and polarizability matrix elements up to 4$^{th}$ order. We also include mechanical anharmonicity and mode-coupling in the model Hamiltonian. This refined RDM model preserves the symmetry of the bromoform molecule.

Lastly, we rigorously account for the inversion symmetry of the liquid by excluding 4$^{th}$ order field interactions from the response. We achieve this by simulating the response of the system to both positive and negative THz fields and combining the two results. The final response is cubic in the field, in agreement with the experimentally measured TTR response. 

The revised experimental spectrum is used to fit the parameters of the updated RDM model, leading to good agreement. The fitted nonlinearities are orders of magnitude smaller than found in our previous work and agree well with \textit{ab initio} electronic structure calculations.  Regardless, the original conclusion that nonlinearities in the dipole surface of the intramolecular vibrations drive the TTR response of bromoform remains unchanged.

\section{Acknowledgements}
The authors thank Ralph Welsch, Ian Finneran, Matthew Welborn, and Philip Shushkov for helpful discussions, as well as Peter Hamm for a sharing a copy of their forthcoming manuscript.  This work was supported by the National Science Foundation (Grant CHE-1665467) and the Office of Naval Research (Grant N00014-10-1-0884).

\bibliography{references}

\begin{figure*}[ht!]
\includegraphics[width=0.99\textwidth]{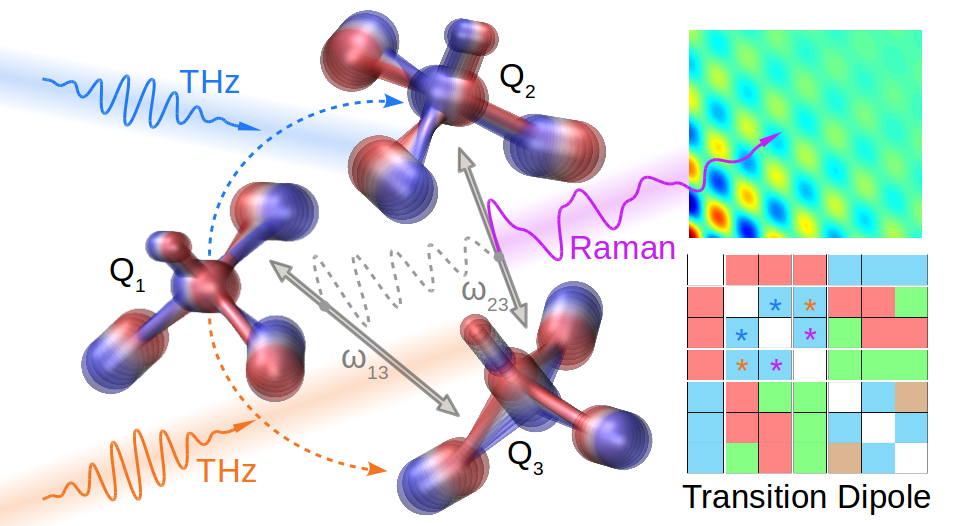}
\caption{TOC figure.}
\end{figure*}

\end{document}